%% file: ms.tex
\documentclass[12pt,preprint]{aastex}

\newcommand{\msun}{M$_{\sun}$}
\newcommand{\kms}{km\,s$^{-1}$}
\newcommand{\ek}{\.E$_k$}

\slugcomment{Running title: Compact sources in M\,82}

\begin{document}


\title{Could the Compact Radio Sources in M\,82 be Cluster Wind Driven Bubbles?}


\author{E. R. Seaquist and M. Stankovi\'c}
\affil{Department of Astronomy\,\&\,Astrophysics, University of Toronto, 50 St. George Street, Toronto, Ontario, M5S 3H4, Canada}

\email{seaquist@astro.utoronto.ca}


\begin{abstract}
The compact non-thermal sources in M\,82 and other starburst galaxies are generally thought to be supernova remnants (SNRs). We consider an alternative hypothesis that most are wind driven bubbles (WDBs) associated with very young super star clusters (SSCs). In this scenario, the synchrotron emitting particles are produced at the site of the shock transition between the cluster wind and the hot bubble gas. The particles radiate in the strong magnetic field produced in the expanding shell of shocked ambient interstellar gas.
One of the motivations for this hypothesis is the lack of observed time variability in most of the sources, implying ages greater than expected for SNRs, but comfortably within the range for WDBs. In addition, as SNRs, these sources are not effective in driving the starburst mass outflow associated with the nuclear region of M\,82, thus requiring a separate mechanism for coupling SN energy to this outflow. 
The WDB hypothesis is found to be feasible for underlying clusters in the mass range $\sim2\times10^{4(\pm1)}\,$\msun\, and ambient gas densities in the range $\sim3\times10^{3(\pm 1)}$\,cm$^{-3}$. The ages of the bubbles are between several $\times10^3$ and several $\times10^4$ years. Since the SNR picture cannont be ruled out, we provide suggestions for specific observational tests which could confirm or rule out the WDB hypothesis.  
Finally, we discuss the WDB hypothesis in the context of broader phenomena in M\,82, such as the rate of star formation and starburst outflows, and the possible interpretation of supershells in M82 as the products of multiple supernovae in young SSCs.
\end{abstract}



\section{INTRODUCTION}

The nearby starburst galaxy M\,82 is host to a population of at least 60 compact (diameter a few pc) radio sources observable at centimeter wavelengths. Similar populations may be found in a number of other starburst systems including the ultra-luminous IR galaxy Arp\,220. For the approximately 60 sources in M\,82, about half are identified (by their spectral indices) as non-thermal or synchrotron emitting sources, and are generally accepted to be the remnants of supernovae associated with individual stars. About a quarter are identified as thermal sources or compact HII regions \citep{mcdonald,rrico} and the rest are not yet identified. The association of the compact non-thermal radio sources (CNRs) with supernova remnants (SNRs) is based on the similarity of their luminosities, morphology and nonthermal spectra to those of SNRs in our Galaxy. 

The purpose of this paper is to explore an alternative interpretation that many, and possibly most, of the CNRs are generated by wind driven bubbles (WDBs) associated with very young super star clusters (SSCs). In this picture, the strong stellar wind from a young cluster energizes a hot expanding bubble which sweeps up interstellar material to form a thin and dense radiative shell \citep[e.g.][]{weaver,koo}. In the suggestion presented in this paper, the non-thermal emission is associated with relativistic electrons accelerated at the adiabatic shock responsible for heating the bubble. The non-thermal emission occurs as these electrons interact with the interstellar magnetic field compressed in the radiative shell comprising the shocked interstellar medium (ISM). Thus the morphology of this WDB would indeed resemble that of an SNR.

But why consider this alternative if the SNR hypothesis works? The motivation is driven by some troubling, if non-fatal, issues with the SNR hypothesis. First, a majority of the sources show no variations in flux density. \citet{kronberg} have studied the time variability of a sample of 24 sources, and have provided a stringent upper limit of 0.1\% per year for the degree of variability in a sub-sample of 18 over 12 years. This suggests a lower limit on the age of at least 1000 years. If, on the other hand, SNRs with diameters up to 4\,pc (comparable to those in M\,82) are in the free expansion stage with velocities $\geqslant$\,5000\,\kms, as suggested by \citet{muxlow}, the corresponding age would be at most a few hundred years. However the SNRs in M\,82 could be decelerated by dense gas, which would make them significantly older. In this case the inferred ages greater than 1000 years correspond to expansion velocities $\lesssim$\,1000\,\kms, and lower than this at later stages. It is difficult to see how such supernovae (SNe) could drive the starburst nuclear outflow at several thousand \kms. Secondly, a significant fraction (perhaps most?) of the massive stellar population is formed in SSCs, as suggested by some authors \citep[e.g.][]{lipscy}. Within such clusters, SNe would be frequent and collectively produce a nearly steady wind rather than individually identifiable SNRs \citep{mccray,maclow}. The result would be shells that look very large and diffuse compared with those for individual SNRs. We return to this point later to discuss evidence for such structures in M\,82.

Some of these difficulties were recently addressed by \citet{chevalier}. To account for the lack of variability, they suggested that most of the SNRs have recently entered the radiative phase as they expand into an ISM of density 10$^3$ H\,atoms\,cm$^{-3}$. The corresponding expansion velocities are $\sim$\,500\,\kms. To drive the starburst wind, they postulate the existence of another population of SNRs expanding into a lower density medium, permitting mechanical energy to be transferred to the starburst wind. Thus the 
SNR model is certainly viable, but in light of the concerns which have spawned such discussion, the WDB hypothesis should be considered.

The WDB hypothesis predicts ages of at least several thousand years for the compact sources, much greater than for SNRs. Figure 1 shows a comparison between the evolution of an SNR (both adiabatic and radiative), and a WDB, both expanding into a medium of density 10$^3$\,cm$^{-3}$. The WDB curves correspond to SSC masses of 10$^4$ and 10$^5$\,\msun, not atypical for M\,82. The WDB evolution represented are described by equations (1) and (2) in \S\,2, where the evolution is discussed in more detail. The point to be made in Figure 1 is simply that for the typical radii of the non-thermal shells (0.5\,--\,2\,pc), WDBs are much older than SNRs. In the WDB picture, the modest shells produced by the stellar winds are replaced after several million years by supershells driven by the collective winds of many SNe within the cluster, as noted earlier. It is the latter winds that are collectively responsible for energizing the starburst outflow.  

\begin{figure}[ht!]
\epsscale{.80}
\plotone{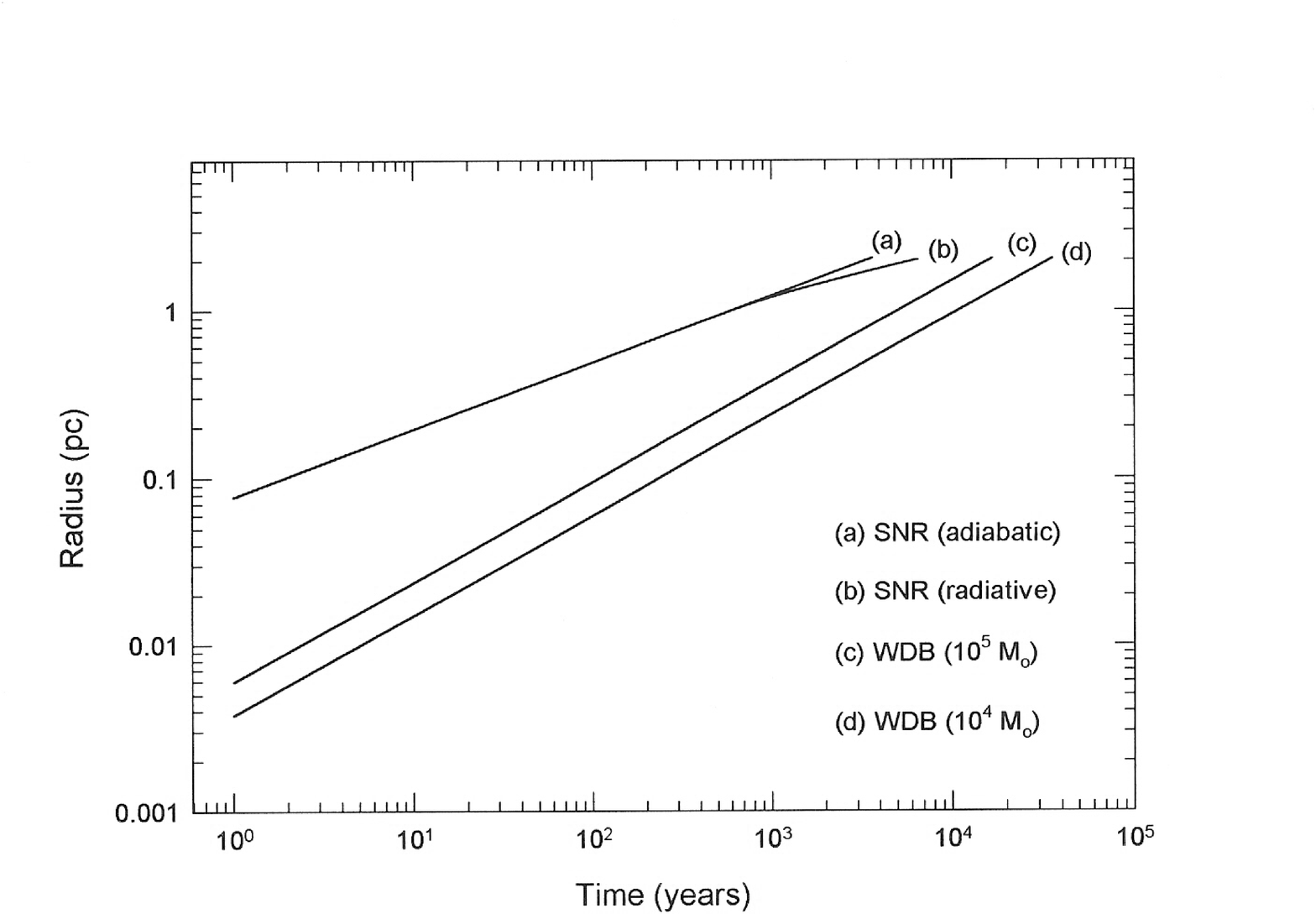}
\caption{Plots of radius vs. time for SNR and WDB shells.  The two curves representing SNRs show adiabatic and radiative behavior, and the two curves representing WDBs are based on winds for cluster masses of 10$^4$ and 10$^5$\,\msun. The ISM density is 10$^3$\,cm$^{-3}$ in all cases.}
\end{figure}

We concern ourselves in this paper with an evaluation of the WDB picture as the explanation for the CNRs, using existing data in the literature. It is not the intention to propose that the WDB hypothesis should now replace the conventional SNR picture, but rather to determine its feasibility by subjecting it to a number of tests. The general approach taken in the paper is to assume that the CNRs have a WDB origin, and then outline and discuss the consequences.  

In \S\,2, we outline the model. Section 3 is devoted to describing tests to determine the feasibility of the model and ranges in parameter space for the viability of the model. The results are presented in \S\,4. In \S\,5, we present inferences from these results concerning the properties of the shells, and in \S\,6 we discuss the relativistic particle acceleration efficiencies with a comparison with Galactic SNRs. Section 7 addresses the issue of the source sizes compared to the radii of the hypothesized underlying clusters, and \S\,8 assesses the likelihood of detecting these clusters and their effects in various wavebands. Finally we present in \S\,9 the conclusions, placing the WDB hypothesis in a larger context of other starburst phenomena in M\,82, and outlining stringent tests for the WDB picture.

\section{OUTLINE OF THE MODEL}

The picture underlying our hypothesis is based on the dynamical evolution of a hot WDB expanding into a uniform surrounding ISM, as developed in a series of papers by \citet{weaver}, \citet{mccray} and \citet{koo}. In this case the wind is the combined winds of all massive stars in a very young SSC, and the ISM is either the ambient density of the molecular cloud containing the SSC, the ambient ISM, or possibly the remnant medium produced by bi-polar flows in the proto-stellar stage of the cluster stars. For simplicity we overlook possible departures in the bubble geometry from that described in the above papers attributable to the finite size of the cluster. The expanding bubble drives a shock into the surrounding ISM forming a thin radiative shell containing swept up interstellar gas and magnetic field. The radius r varies according to the similarity solution,

\begin{equation}
{\rm r=0.0170\,(L_{38}/n_0)^{1/5}\,t^{3/5}\,pc}
\end{equation}

where t is the time (yr), L$_{38}$ is the wind power in units of 10$^{38}$\,erg\,s$^{-1}$, and n$_0$ is the uniform ISM density in H\,atoms\,cm$^{-3}$. This equation is analogous to the Sedov adiabatic blast wave similarity solution relevant to an adiabatically expanding SNR, given by

\begin{equation}
{\rm r=0.310\,(E_{51}/n_0)^{1/5}\,t^{2/5}\,pc }
\end{equation}

where E$_{51}$ is the mechanical energy associated with the blast ejecta in units of 10$^{51}$\,erg.

Equations (1) and (2) are represented in Figure 1, except for the addition of a curve for the radiative phase of SNR expansion which simply conserves momentum. For the SNR behavior we have used ${\rm E_{51}=1}$ and ${\rm n_0=10^3}$\,cm$^{-3}$. The radiative phase is assumed to begin at ${\rm r=1}$\,pc, appropriate for an ISM density of 10$^3$\,cm$^{-3}$ \citep[see for e.g.][]{chevalier}. The wind power levels in the figure are based on models from Starburst99 (hereafter SB99) \citep{leitherer}. These levels are presented in \S\,3.   

The interior of the wind bubble comprises a rarified hot (${\rm T\sim10^6}$\,K) plasma, the consequence of a shock transition between the hypersonic stellar wind and the gas in the bubble cavity (see also section 6.3). The hot bubble is surrounded by a contact discontinuity separating the bubble from the shocked ISM. In this model the bubble shock is the source of the relativistic electrons causing the observed synchrotron radiation. The relativistic particles are assumed to propagate toward the contact discontinuity given by equation (1), where they diffuse into the region of the radiative dense shell and emit synchrotron radiation in the compressed magnetic field of the ISM. In addition, these electrons may be further accelerated at the interstellar shock. 

The mechanism for accelerating the relativistic particles is diffusive shock acceleration, or first order Fermi mechanism, involving particles trapped between collapsing mirrors formed by the bubble shock and upstream magnetic irregularities \citep[see for e.g. reviews by][]{drury,blandford}. The mechanism is identical to that thought to produce the synchrotron emitting electrons in SNRs and many other relevant astrophysical scenarios. We note in particular the shock induced synchrotron emission in Wolf-Rayet binaries \citep[e.g.][]{pittard} and shocks embedded in O star winds \citep{vanloo}. One of the scenarios described in the aforementioned reviews is particle acceleration at the termination shock of a stellar wind, which is essentially the nature of the bubble shock. The treatment of the acceleration process is beyond the scope of this paper. Instead, the production of non-thermal electrons involving shock speeds of several thousand \kms\, is taken as assured, since it occurs for example in SNRs. The issue then focuses on the acceleration efficiency, which is discussed in \S\,6.

The model implicitly assumes that the accelerated particles reach the outer emitting region in a time short compared with the age of the shell. Though the particles will tend to stream outward from the acceleration zone at relativistic speeds, they would undoubtedly be scattered by magnetic irregularities in the hot bubble. The material in the bubble is composed primarily of gas (with embedded fields) evaporated from the cool outer shell \citep{weaver}. The treatment of this problem is outside the scope of this preliminary paper, but we note that the magnetic irregularities may take the form of Alfv\`en waves produced by gas turbulence or generated by the collective behavior of the streaming energetic particles themselves \citep[e.g.][]{wentzel}. Under these conditions, the limiting speed of propagation of the relativistic particles is the Alfv\`en speed of the waves. If the energy density associated with these waves is comparable to the thermal energy density in the hot bubble, then the Alfv\`en speed is comparable to the sound speed (and thus the speed of the stellar wind). Such conditions would lead to a propagation time which is short compared to the age of the WDB.

\section{FEASIBILITY TESTS}

The model outlined in \S\,2 was applied to published radio data for a sample of well observed CNRs in M\,82 to determine for each source its feasibility and resulting constraints on SSC mass and ISM density (see \S\,4). This sample is hereafter referred to as the input sample for the array of tests. The output sample comprises {\it all} sources from the input sample which pass all of the tests, yielding for each member a range of SSC mass and ISM density. The members of the output sample are shown in Table 1. We outline below the tests which, by their stringency, act to most severely test and constrain the model. Some of the tests invoke physical plausibility (e.g.\, \S\,3.1), whereas others refer to limits set by observations (e.g.\, \S\,3.5). Both forms of test combine to impose limits on the cluster mass and/or the ambient gas density surrounding the bubble. Some further less stringent constraints, additional insights, and further discussion of plausibility are deferred to \S\,5\,--\,8. The sole selection criterion for membership in the input sample is the requirement for observed flux densities at a minimum of four wavelengths, and thus well defined non-thermal spectral indices. The sample was drawn from data in \citet{huang}, \citet{muxlow}, \citet{allen}, and \citet{mcdonald}. Excluding the strongly variable candidate 41.9\,+\,57.5, the input sample comprised 27 sources. Note that the input source sample contains a number of sources whose measured flux variability \citep{kronberg} and/or expansion rates \citep{muxlow05,fenech,beswick} are consistent with or suggest the SNR hypothesis. If such sources pass the WDB feasibility tests, then the tests do not discriminate between the models.
The tests applied are outlined individually in \S\,3.1 through 3.5.

\subsection{Energy conservation}

Conservation of energy requires the wind to supply sufficient mechanical energy over its age for the requirements of the synchrotron sources. The relativistic particle and magnetic field energies were derived from equipartition arguments in \citet{pacholczyk}, which provide minimum energy requirements. 
The required flux densities, spectral indices, and angular sizes for individual sources were obtained from \citet{huang}, \citet{muxlow}, \citet{allen}, and \citet{mcdonald}. We assumed spherical geometry with source diameters equal to the average measured along major and minor axes. For a few sources without published angular sizes we adopted an angular diameter of 0.2\arcsec, following \citet{allen}. Radio luminosities were computed over the standard frequency range 0.01 to 100\,GHz, and the distance adopted for M82 was 3.25\,Mpc.
 The total equipartition energy may be expressed in the form ${\rm E_p=E_1\,(1+k)^{4/7}\,\Phi^{3/7}}$\,erg, where k and $\Phi$ are respectively the (unknown) ratio of proton to electron energy and volume filling factor for the synchrotron source. E$_1$ is then the minimum energy requirement when ${\rm (1+k)^{4/7}\,\Phi^{3/7}=1}$. Note that the equipartition argument requires that the energy be divided approximately equally between relativistic particles and magnetic field energy. The quantity E$_1$ was computed for each source, resulting in a median value for the output sample of $3.8\,\times10^{48}$\,erg and a range from $1.4\,\times10^{48}$ to $7.1\,\times10^{48}$\,erg. The result for each source in the output sample is shown in Table 1, which also shows the equipartition magnetic field ${\rm H_1\times [(1+k)/\phi]^{2/7}}$ with the coefficient of H$_1$ set to unity.

The mass and energy outflow rates of the cluster wind are assumed to be those associated with the beginning of the main sequence phase. It is possible that outflows could be associated with massive protostars, but the outflow behavior for this phase is not well understood or presently computable. The wind parameters were derived using SB99 models assuming a cluster with an IMF with low and upper mass cutoffs of 0.1 and 100\,\msun\, respectively and a break in slope at 0.5\,\msun. Above the break, the slope is $-$2.30, and below the break, the slope is $-$1.30. The derived values for the wind and luminosity, with their dependence on SSC mass M$_4$ (in units of 10$^4$\,\msun\,) are then:
\begin{mathletters}
\begin{eqnarray}
{\rm {Mass\,loss\,rate\,(dM/dt)}} &  = & 1.86\,\times\,10^{-5}\,{\rm M_4\,M_{\sun}\,yr^{-1}} \\
{\rm Wind\,power\,(L_w)} &  = & 5.61\,\times\,10^{37}\,{\rm M_4\,erg\,s^{-1}} \\
{\rm Wind\,velocity\,(V_w)} & = & 3000\,{\rm km\,s^{-1}} \\
{\rm Bolometric\,luminosity (L_{bol})} & = & 3.40\,\times\,10^{40}\,{\rm M_4\,erg\,s^{-1}} \\
{\rm Lyman\, continuum\, luminosity (L_y)} & = & 4.14\,\times\,10^{50}\,{\rm M_4\,photons\,s^{-1}} 
\end{eqnarray}
\end{mathletters}
The requirement adopted for energy conservation is that ${\rm E_1/E_w = f\,(\leqslant\,1)}$, where \mbox{${\rm E_w = L_w\,t}$}, and t is the age of the wind. The age t is determined for any value of M$_4$ and n$_0$ by equations (1) and (3b). The parameter f thus expresses the efficiency of conversion of wind energy to relativistic particle and magnetic field energy. The tests were made using f = 0.1, corresponding to a conversion efficiency into particle and magnetic field energy of 10\% if the factor (1+k)$^{4/7}\Phi^{3/7}$ = 1.0. The choice f = 0.1 is discussed and justified in \S\,6.1.

\subsection{Radiative energy losses by relativistic electrons}

The relativistic electrons are subject to losses by synchrotron radiation and inverse Compton scattering. The Compton scattered photons in this case are from the cluster stars. The time scale or half-life of these two effects for electrons radiating at $\nu$ (GHz) are given by
\begin{mathletters}
\begin{eqnarray}
{\rm T_{SYNC}} & = & 0.61\,{\rm H^{-3/2}\,\nu^{-1/2}\,   yr}		\\					
{\rm T_{IC}} & = & 1.74\,\times\,10^7\,{\rm H^{1/2}\,r^2\,L_{40}^{-1}\,\nu^{-1/2}\,   yr}.			
\end{eqnarray}
\end{mathletters}
Here H is the magnetic field strength (gauss), r is the radius of the emitting shell (pc), and L$_{40}$ is the bolometric luminosity of the cluster (10$^{40}$\,erg\,s$^{-1}$). The test requires that ${\rm T_{SYNC}, T_{IC} > t}$ for $\nu$ = 5\,GHz to avoid significant depletion of particle energy in the observed region of the radio spectrum. Such depletion would produce a break in the radio spectrum at 5\,GHz, which is not observed. The magnetic field adopted is the equipartition field that depends on the unknown parameters (k, $\Phi$) as [(1+k)/$\Phi$]$^{2/7}$. This factor was set to unity in the computation of the quantity H used in equations (4a) and (4b). Note that the dependence on this unknown factor is rather weak, especially in equation (4b), where ${\rm T_{IC}}\sim[(1+k)/\Phi]^{1/7}$. 

We note that other types of particle losses may occur, including expansion and ionization losses, but these are less certain, and consideration of these is deferred to \S\,6.3.

\subsection{Radiative losses by the hot bubble}

The initial expansion of the bubble is adiabatic, expanding according to equation (1), but when the radiative phase sets in, the bubble pressure drops and no longer drives the shock into the ISM. The expansion rate of the shell then slows significantly following an expansion law of ${\rm r\sim t^{1/4}}$ until the subsequent wind momentum exceeds the present momentum in the shell. Radiative losses are important when ${\rm t = T_{COOL}}$, defined as the time when the total energy radiated by the bubble is equal to the total energy input from the wind. The cooling time ${\rm T_{COOL}}$ is given by the following equation based on that given by \citet{mccray} (for solar metallicity),
\begin{equation}
{\rm T_{COOL} = 7.3\,\times\,10^{-5}\,L_w^{0.3}\,n_0^{-0.7}\,   yr}
\end{equation} 
An additional effect of bubble cooling is that the acceleration of relativistic particles will shut off at ${\rm t = T_{COOL}}$, since the bubble shock collapses to the outer radiative shell where the higher gas densities preclude efficient particle acceleration. The low acceleration efficiency is a consequence of excessive collisional losses at the low energy phase of the acceleration process. Accordingly, this test requires that the bubble be in the adiabatic phase, i.e. ${\rm T_{COOL} > t}$.

\subsection{Pressure confinement}

The expanding shell is subject to confinement due to the outside ambient gas and magnetic pressures of the ISM. This pressure stops the expansion when it becomes 
comparable to the ram pressure of the expanding shell \citep[e.g.][]{oey}. We define
the confinement time ${\rm T_{CONF}}$ to be the time at which the shell expansion slows to 10\,\kms, adopted as the effective sound speed of the ISM in M\,82. We thus require
${\rm T_{CONF} > t}$, where the expansion rate is given by dr/dt = $\frac{3}{5}$\,r/t.

\subsection{Presence of a detectable HII region surrounding the cluster}

The most massive stars in the cluster will produce a surrounding HII region which should be detectable as a thermal radio source. Such emission is probably the source of the compact HII regions seen in M\,82, which so far have no detectable non-thermal emission. Since the expanding shell should be either ionization bounded, or nearly so (see \S\,5.2.3), the expected thermal or free-free flux density ${\rm S_{EXP}}$ is derivable directly from the Lyman continuum luminosity Ly given by equation (3e). The expected (optically thin) radio flux is,
\begin{equation}
{\rm S_{EXP} = 1.32\,\nu^{-0.1}\,Ly_{50}\,D^{-2}\,   mJy},
\end{equation}
where ${\rm Ly_{50}}$ is in units of 10$^{50}$\,photons\,s$^{-1}$, and D is the distance in Mpc (in this case 3.25\,Mpc). This emission is not significantly attenuated by dust, and is characterized by its flat or rising radio spectrum, compared to the non-thermal emission which has a negative slope.

The difficulty with applying this test to the SNR sources is the confusion by non-thermal emission. Since the spectrum of optically thin free-free emission is flat, detectable thermal emission would manifest itself as a positive curvature of the radio spectrum, affecting the spectrum preferentially at high frequency (in this case 15\,--\,20\,GHz). In order to search for a thermal component, the radio spectrum for each member of the input sample was fit by a curve representing combined thermal and non-thermal emission. In no case was there evidence for a thermal component. A 3$\sigma$ upper limit ${\rm S_{OBS}}$ was therefore assigned in each case, where $\sigma$ is the standard error of the thermal component derived from the fit. Thus the test in this case requires that the expected flux not exceed the observed upper limit, i.e. ${\rm S_{EXP} < S_{OBS}}$, where the typical value for ${\rm S_{OBS}}$ is a few tenths of a mJy.

\section{RESULTS}

In applying each of the limits above, a violation by a factor of two was permitted to allow for the uncertainty in deriving the physical parameters. For example, the energy test in \S\,3.1 was regarded as successful as long as the energy requirement of relativistic electrons did not require the parameter f to exceed the prescribed value of 0.1 by more than a factor of two.

For each source in the input sample, the five tests were applied considering ranges in possible SSC mass and ISM density of $10^3-10^6$\,\msun\, and $10-10^5$\,cm$^{-3}$, respectively. The logarithmic step size was 0.5 for both parameters.
The result is an output sample comprising sources which passed all five tests described in \S\,3 for a specified range in (M$_4$, n$_0$).
Briefly stated, the mass range is constrained by tests 3.1 and 3.5, and for a given mass (i.e.\,wind power), the ISM density is constrained on the low side by ages too low to provide sufficient energy in test 3.1, and by tests 3.2 and 3.3 on the high side by excessive radiation or cooling losses. Test 3.4 did not provide any limiting constraint within the ranges allowed by the other tests, signifying that the shells are not pressure confined. Table 1 lists the members of the output sample with the inferred medians and ranges of SSC mass and ISM density. 
The output sample contains all but two members of the input sample, namely 42.5\,+\,61.9 and 42.8\,+\,61.3, both of which failed test 3.5. Failure of this test signifies that the masses (and hence Lyman continuum luminosities) required were so high that the predicted free-free emission from the associated HII regions was higher than the upper limits set by observations. Thus it may be concluded that the WDB model is consistent with the great majority of CNRs, though this does not preclude their interpretation as SNRs. It remains distinctly possible as well that CNRs comprise a mixture of both types.

Figure 2 shows the distributions of the median values in Table 1. The distribution shows no significant correlation between M$_4$ and n$_0$, and the centroid of this distribution corresponds to ${\rm M = 1.8\times10^4}$\,\msun\, and ${\rm n_0 = 2.6\times10^3\,cm^{-3}}$. Since the mass range for observed SSCs is about 10$^{5(\pm\,1)}$ \citep{melo}, we conclude that if the CNRs are WDBs, the SSCs responsible are at the low end of the mass range for M\,82. The derived value of n$_0$ is consistent with the density associated with the inter-clump medium of M\,82 ($\sim$\,10$^3$\,cm$^{-3}$) rather than the molecular cloud densities ($\sim$\,10$^5$\,cm$^{-3}$) \citep[see for e.g.][]{gusten,mao}, which suggests that the clusters would have consumed most of the dense molecular gas in their parent clouds. 

It must be emphasized that the tests do not lead to a discrimination between the WDB and SNR models. In particular, some sources in the output sample exhibit variability consistent with the SNR hypothesis, and a small number have measured expansion velocities of several thousand \kms\, suggesting an SNR in free expansion. However, it is interesting that at least two of sources in the latter category show no measurable decrease in flux as would be expected for a freely expanding cloud of particles and magnetic field. \citet{fenech} report an expansion velocity of 10,500\,\kms\, for 43.2\,+\,58.4, based on an apparent increase in size of 0.7\,mas per year between 1992 and 2002. This figure corresponds to an increase of about 1\%\, per year in radius. However, a linear least squares fit to the data for radio flux vs.\,data in \citet{kronberg} yields a change in flux of $+0.50\pm0.54$\%\, per year over the period 1982--1992. Similarly, \citet{beswick} derive an expansion velocity near 10,000\,\kms\, for the source 43.3\,+\,59.2. Their Figure 8 shows that the rate of increase in source size, extrapolated back to 1982--1992 yields at least 4\%\, per year, whereas the flux data yield a change of $+0.40\pm0.22$\%\, per year over this period. These apparent inconsistencies suggest that it may yet be premature to use the measured expansion velocities to conclude anything definitive about the nature of these sources.

\begin{figure}[ht!]
\epsscale{.80}
\plotone{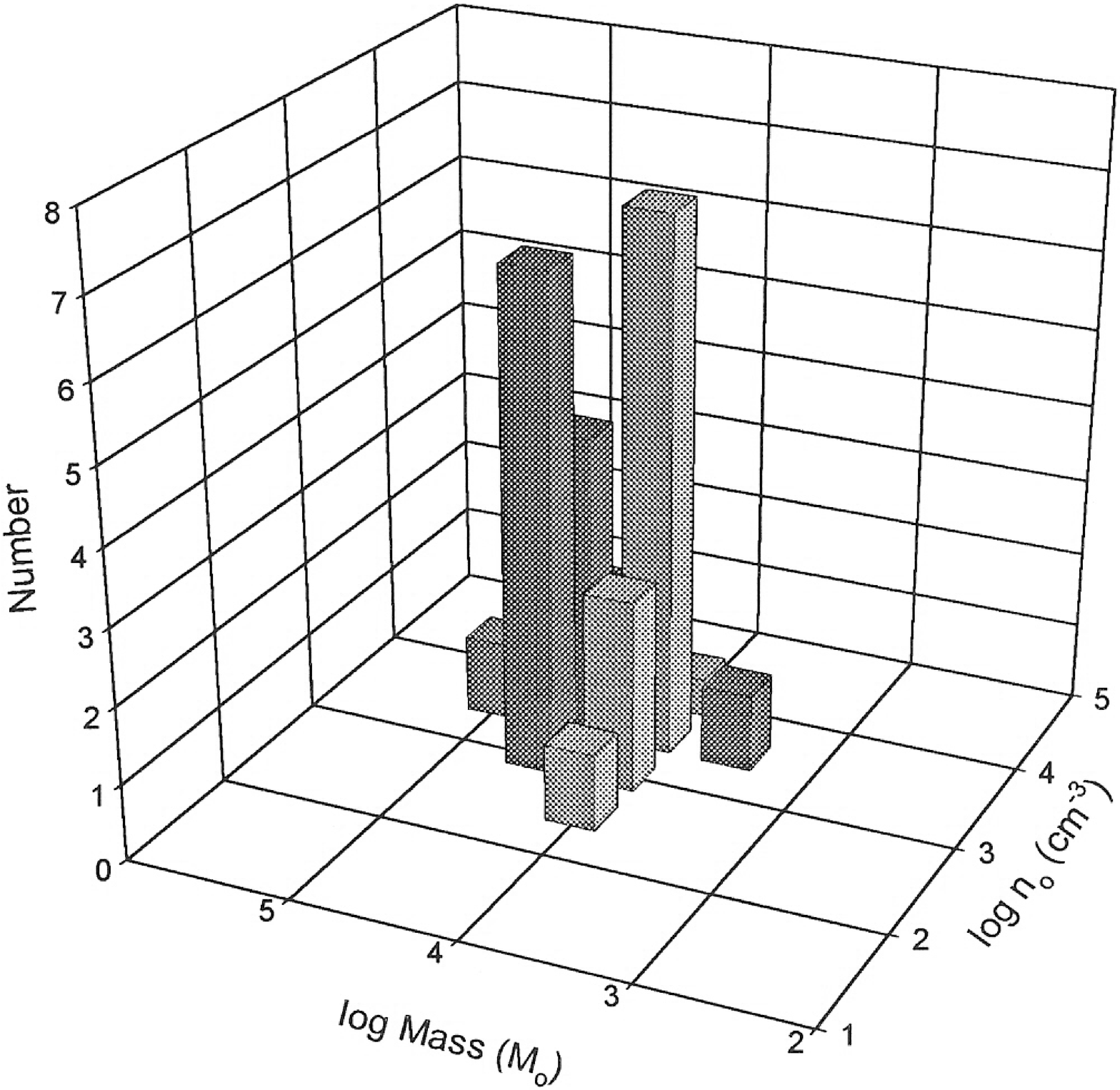}
\caption{Plot in 3-D showing the range in parameter space for cluster mass and ambient ISM density possible for a WDB interpretation of a representative sample of compact non-thermal radio sources. The individual values are medians of the range of acceptable values for each source (see Table 1).
}
\end{figure}

\section{DISCUSSION}

\subsection{Possible similar origin between CNRs and compact HII sources}

Though there is no known relationship between the non-thermal and thermal compact radio sources (i.e.\,compact HII regions), it is plausible that both may be associated with SSCs. The median radius of the compact HII regions measured by \citet{rrico} is about 4\,pc, comparable to and somewhat larger than those for the CNR population (1\,pc). The HII regions may be excited by massive stars in the clusters, which should also produce WDBs. Proceeding on the assumption that the compact HII regions are associated with SSCs, we used their 43\,GHz flux densities measured by \citet{rrico}, together with equations (3e) and (6), to derive the masses of the associated clusters, assuming that they have the same IMF as those associated with the CNRs. Figures 3(a) and 3(b) show a comparison between the inferred distribution of cluster masses underlying CNRs (from Figure 2), and the corresponding distribution for the HII sources. The compact HII regions are seen to be associated with significantly more massive SSCs. The median mass for the HII region clusters is $1.1\,\times\,10^5$\,\msun\, compared to $1.8\,\times\,10^4$\,\msun\, for the CNR clusters. By the WDB hypothesis then, the compact HII regions would be associated with young clusters of mass typical for M\,82, while the CNR clusters would be under-massive.

\begin{figure}[ht!]
\epsscale{.80}
\plotone{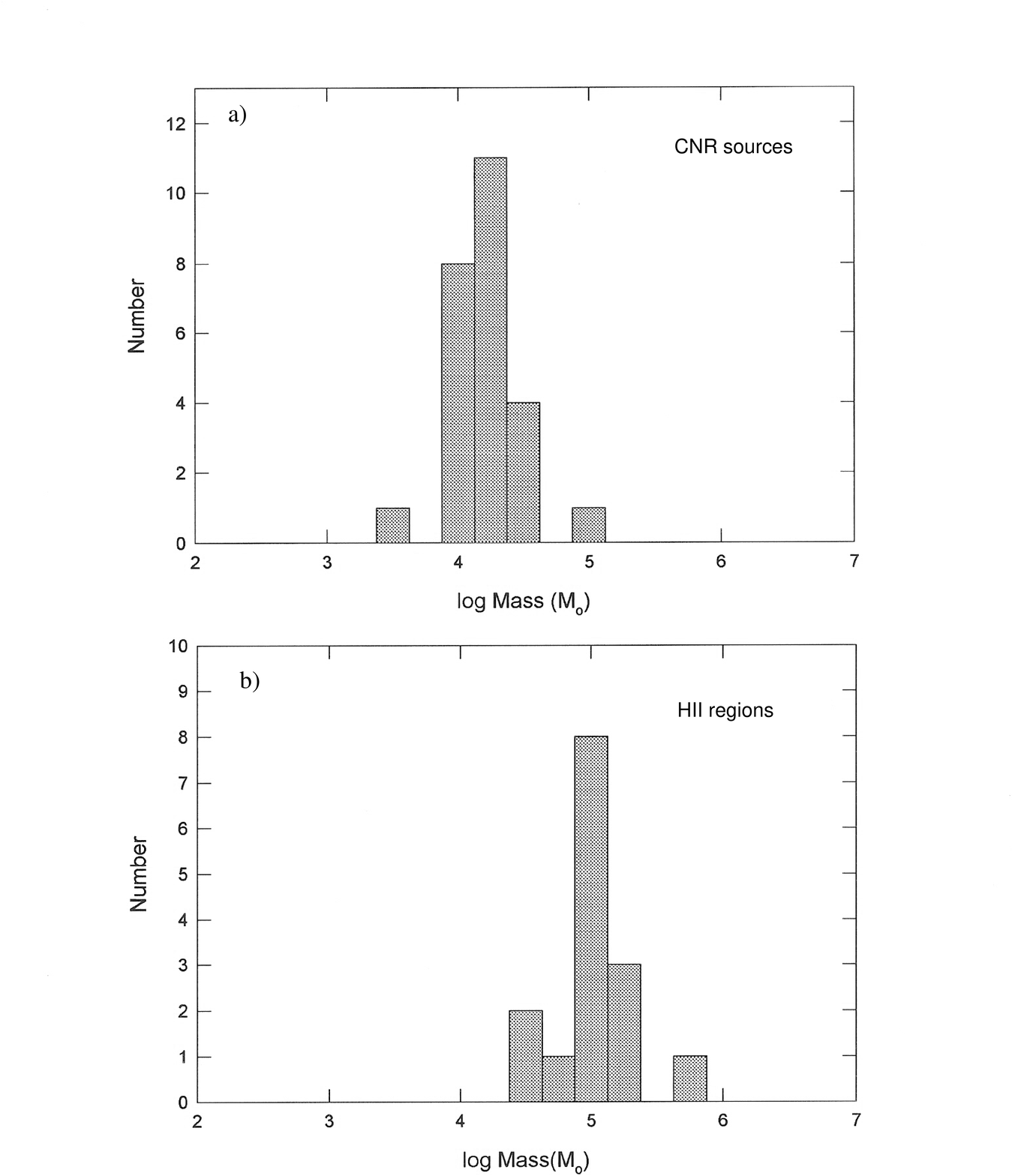}
\caption{Distribution of (a) allowed masses for the WDB model of the compact non-thermal sources, based on Figure 2 and (b) the same for compact HII regions based on the observed 43\,GHz flux densities by \citet{rrico}.
}
\end{figure}

If compact HII regions are associated with cluster WDBs, and if WDBs invariably produce synchrotron emitting electrons, then the lack of detectable non-thermal emission from the HII sources is puzzling. The absence is possibly attributable to shorter inverse Compton lifetimes associated with the higher cluster luminosities. These bubbles may also be older, in their radiative phase, and/or pressure confined. In this case particle acceleration will have ceased. 

In order to identify the most likely possibilites, we adopt ${\rm M=1.1\,\times\,10^5}$\,\msun\, for the underlying clusters, ${\rm r=4}$\,pc for the associated bubble radius, as described earlier, and ${\rm H=10^{-3}}$\,gauss for the emitting region. We applied the same tests to this case as for the CNRs, except for test 3.5, since the free-free flux in this case was used to estimate the cluster mass. We find that nonthermal emission is possible provided ${\rm n_0\lesssim\,10^4}$\,cm$^{-3}$, for which the bubble age inferred from equations (1) and (3b) is less than about 10$^5$\,yr. At higher density the corresponding age would be greater than the inverse Compton lifetime for electrons emitting at 5\,GHz. At ${\rm n_0\approx\,10^5}$\,cm$^{-3}$ radiative cooling also becomes an important factor, shutting off the supply of relativistic particles. The conclusion is that synchrotron emission would be supressed in the observed HII sources if the ambient gas density exceeds 10$^4$\,cm$^{-3}$.

\subsection{Properties of the WDB shells for the CNRs}

\subsubsection{Mass, column density and age}

Figures 4, 5, and 6 show the distributions of the inferred mass, column density and age for the shells in the WDB hypothesis. These parameters were derived from the median values of (M, n$_0$) for each member of the output sample in Table 1. The shell masses are derived from the source volume and the ISM density. The column density of hydrogen ${\rm N_S}$ associated with the shell thickness is derived using ${\rm N_S = 1/3\,n_0\,r}$, appropriate for a thin shell, and the ages are derived from equation (1). 

\input{tab1}

Note that the shapes of the distributions reflect largely the uncertainties in the determination of these parameters rather than their intrinsic distributions. It is the median values that are the most relevant. The median shell mass and column density are 212\,\msun\, and $2.3\,\times\,10^{21}$\,cm$^{-2}$, respectively. Using a figure of $1.87\,\times\,10^{21}$\,cm$^{-2}$ for a visual extinction of 1 magnitude \citep{bohlin}, the median shell extinction is 1.2$^m$, indicating that many of the shells in this picture are partially transparent to optical radiation. Thus any neutral gas would be significantly photo-dissociated, comprising mainly HI instead of H$_2$. The median age of the winds is about $1.3\,\times\,10^4$ yr, significantly greater than for the SNR hypothesis ($\leqslant$\,10$^3$\,yr). The derived ages have implications for the star formation rate in clusters with masses near 10$^4$\,\msun. From the median mass of $\sim2\,\times\,10^4$\,\msun\, for the 25 SSCs in the output sample and adopting a maximum age of $\sim5\,\times\,10^4$\,yr, the implied SFR is $\sim10$\,\msun\,yr$^{-1}$. This value seems not unreasonable compared to recent estimates for the total SFR of up to 10\,--\,33\,\msun\,yr$^{-1}$ \citep{forster}.

\begin{figure}[ht!]
\epsscale{.80}
\plotone{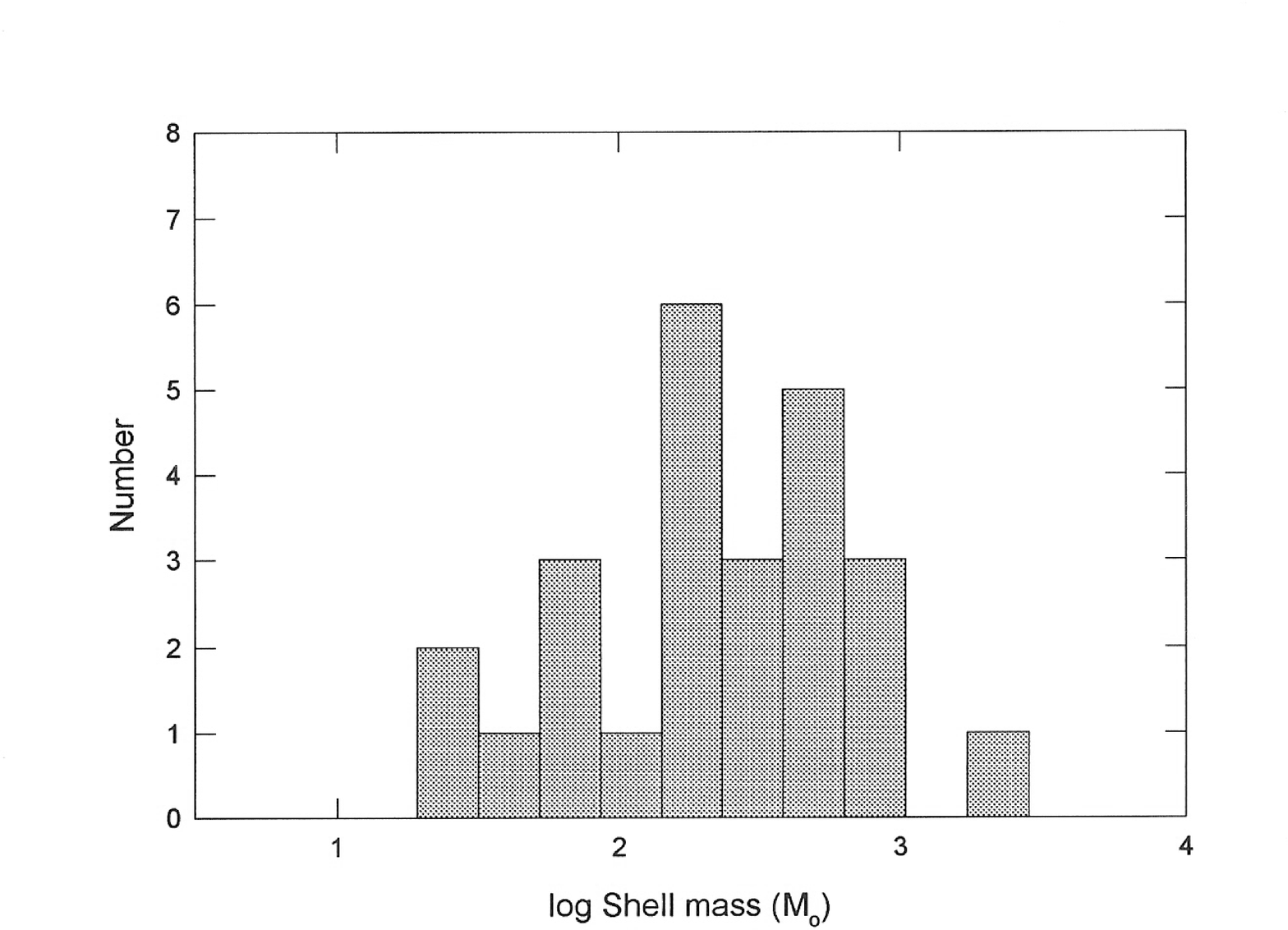}
\caption{The distribution of $\log$\,Mass contained in the shells swept up by the expanding bubble in the WDB interpretation of compact non-thermal sources, computed from the size and median gas density associated with each source in Table 1.}
\end{figure}

\begin{figure}[ht!]
\epsscale{.80}
\plotone{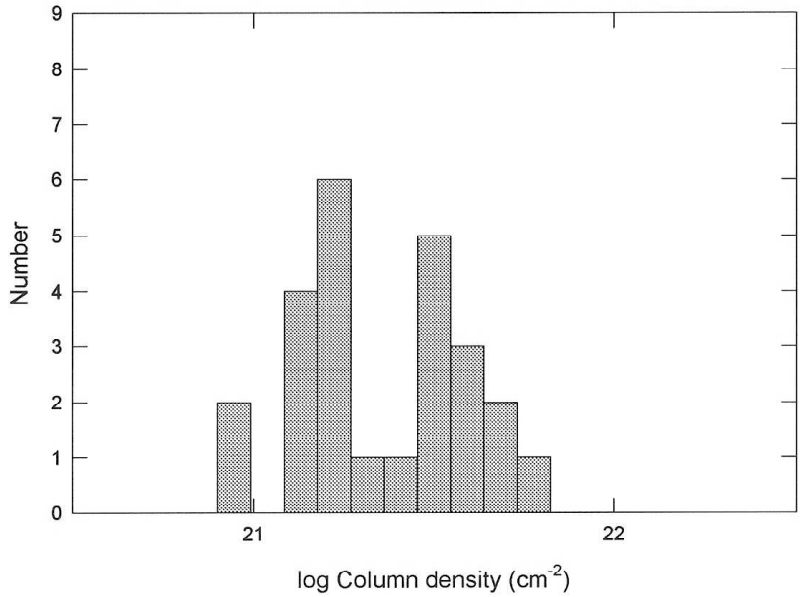}
\caption{The distribution of $\log$\,Column Density associated with the shells swept up by the expanding bubble in the WDB interpretation of compact non-thermal sources, computed from the size and median gas density associated with each source in Table 1.}
\end{figure}

\begin{figure}[ht!]
\epsscale{.80}
\plotone{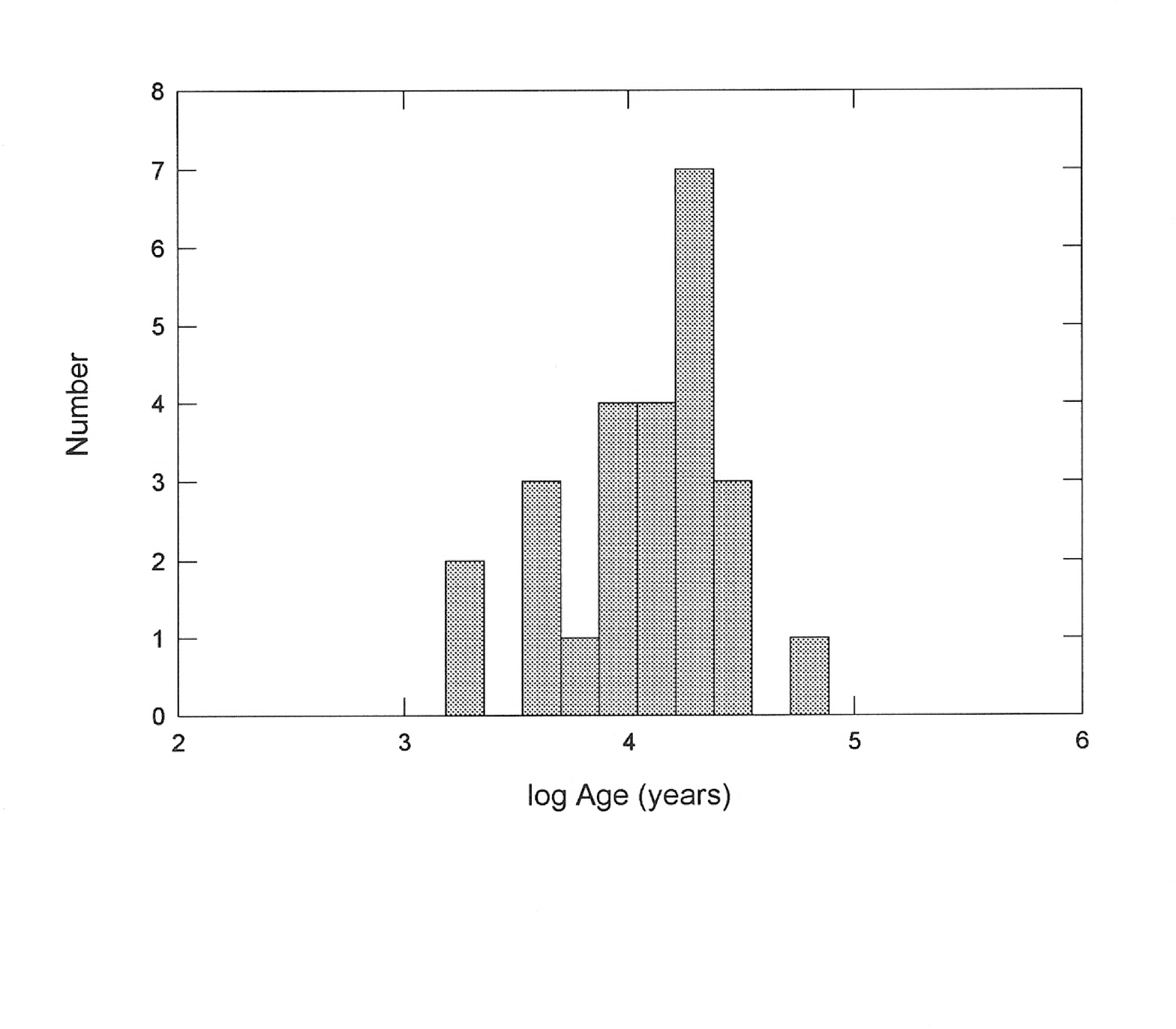}
\caption{The distribution of $\log$\,Age associated with the shells swept up by the expanding bubble in the WDB interpretation of compact non-thermal sources, computed from the median mass (and associated wind power) and median gas density associated with each source in Table 1.}
\end{figure}

\subsubsection{Magnetic field and gas density}

The magnetic field is not directly observable, but is assumed to be the equipartition field derived from the synchrotron luminosity. The median value is ${\rm 5.7\,\times\,10^{-4}\,(1+k)^{2/7}\Phi^{-2/7}}$\, gauss.  However, the WDB model provides an independent (though model dependent) method for estimating the magnetic field in the expanding shell using pressure balance between the field in the shell and the ram pressure of the ISM. Assuming for the moment that gas pressure in the shell is not a significant contributor, and that the magnetic field is the only means of support, the field strength is given by ${\rm H^2/8\,\pi \,\approx\,\rho\,V_S^2}$ where ${\rm V_S}$ is the shell expansion velocity. Taking ${\rm V_S =\frac{3}{5}\,r/t}$, the median shell speed for the sources is $\sim$56\,\kms, and the corresponding value for the magnetic field in the shell is $1.5\,\times\,10^{-3}$\,gauss. This figure is close to median equipartition field if allowance is made for the factor (1+k)$^{2/7}\Phi^{-2/7}$ which will exceed unity. It is interesting to compare this value with the ambient interstellar field in M\,82, which is between $5\,\times\,10^{-5}$ and $1.4\,\times\,10^{-4}$\,gauss \citep{rieke,klein,condon,thompson}. Assuming a value of 10$^{-4}$\,gauss for the interstellar field and adopting $1.5\,\times\,10^{-3}$\,gauss for the field in the shell, we find that interstellar field is compressed by about one order of magnitude. Assuming the compression is 1-D with the field parallel to the shock, and that the field is frozen in, the shell gas density n$_{\rm s}$ corresponding to the median value of ${\rm n_0 = 2.6\,\times\,10^3}$\,cm$^{-3}$ is ${\rm n_s = 3.9\,\times\,10^4}$\,cm$^{-3}$. Two factors which could act to lower this density are (1) a lower value for the shell magnetic field by taking account of other possible sources of pressure, and (2) a higher value for the interstellar magnetic field, i.e.\, stronger than the equipartition value, as suggested recently by \citet{thompson}.
Since this density cannot provide significant pressure at the low temperatures expected for the radiative shell, the initial assumption of ignoring gas pressure appears justified.

\subsubsection{Fractional ionization}
 
The hydrogen emission measure (${\rm EM_B}$) of a geometrically thin ionization bounded shell can be determined directly from the cluster Lyman continuum luminosity, which when combined with equation (3e), yields
\begin{equation}
{\rm EM_B = 4.3\,\times\,10^6\,M_4 / r^2\, cm^{-6}\, pc}
\end{equation}
This equation assumes an ionized gas kinetic temperature of 10$^4$\,K and Menzel case B recombination. For the median cluster mass of $1.8\,\times\,10^4$\,\msun, equation (7) yields ${\rm EM_B = 7.7\,\times\,10^6}$\, cm$^{-6}$\,pc, using a mean radius of 1.0\,pc. This value may be compared with that occurring if the shell were fully ionized, which is,
\begin{equation}
{\rm EM_S  = 3.2\,\times\,10^{-19}\,n_s\,N_S\,cm^{-6}\,pc},
\end{equation}
where ${\rm N_S}$ is the column density in cm$^{-2}$. With ${\rm n_s = 3.9\,\times\,10^4}$\,cm$^{-3}$ and ${\rm N_S = 2.3\,\times\,10^{21}}$\,cm$^{-2}$ (see \S\,5.2.1 and 5.2.2), equation (8) yields ${\rm EM_S = 2.9\,\times\,10^7}$\,cm$^{-6}$\,pc. Thus the ratio ${\rm EM_B/EM_S}$\,$\sim\,0.3$, indicating that a typical shell is probably ionization bounded. However, the ratio is so close to unity that some shells would be substantially neutral and others substantially ionized.

\section{ENERGETICS FOR WDB AND SNR MODELS}

The radio luminosities of the CNRs are higher than for galactic SNRs. The total kinetic energy for SNR ejecta is 10$^{51}$\,erg, whereas the total energy available in the WDB model for the median mass and source age is about $4\,\times\,10^{49}$\,erg. Thus at first sight, the efficiency of relativistic particle acceleration in the WDB interpretation needs to be at least an order of magnitude higher than for the SNR interpretation. We examine this argument in more detail below.

\subsection{Relativistic particle efficiencies in Galactic SNRs}

Some estimates of the efficiency of relativistic particle acceleration $\eta$ in SNRs are based on a comparison between the total equipartition energy for the relativistic particles and the total kinetic energy of 10$^{51}$\,erg for the ejecta. An early estimate of this type by \citet{ginzburg} led to $\eta =1-10\%$, but with very large uncertainties. A primary difficulty with this method lies in the uncertain ratio k of energy between protons (which radiate inefficiently) and electrons (which provide essentially all of the observed synchrotron emission). Another method is based on direct estimates of the local cosmic ray energy density and the assumption that these are generated solely by SNRs. For this method, estimates of the cosmic ray residence time in the disk and the rate of Galactic SNe are required. These estimates vary widely because of the uncertainty in the parameters used. Using the latter method, \citet{blandford} find $\eta = 3\%$. At the other extreme, \citet{fields}, using the same method, find $\eta = 30\%$, admitting an uncertainty of a factor of at least three. Theoretical estimates also tend to be rather high, with figures that reach 50\% \citep{berezhko}. Note that in local cosmic rays the fraction of electrons is only $\sim$\,2\%, leading to corresponding electron acceleration efficiencies ${\rm \eta_e}$ ranging from $\sim$\,0.06\% to $\sim$\,1\%. 

Recent studies of TeV gamma ray emission produced by inverse Compton scattering of cosmic background photons by energetic electrons provide fresh insight into the efficiency of electron acceleration, concluding that in some SNRs $\eta_{\rm e}$ is in the range 1.0\,--\,2.5\% \citep[and references therein]{keshet}. The corresponding fraction of shock thermal energy carried by relativistic electrons is then about 5\%. Note that the proton/electron energy ratio k must then be much lower than 50, and perhaps nearer 10 to keep the proton contribution from dominating the thermal gas pressure. \citet{ellison} estimates that the fraction of post shock thermal energy occupied by relativistic particles (including ions) in SNR shocks is about 50\%. If we regard this as a reasonable upper limit, then combining this with the figure of 5\% for the partial pressure of electrons, yields k\,$\leqslant$\,10.  In order to be compatible with the observed radio luminosities of Galactic SNRs, the large fraction of shock thermal energy in the form of relativistic particles implies a high ratio of particle/magnetic field energy, so that at least some SNRs are far from equipartition \citep[e.g.][]{dyer}. 

Presuming that the non-thermal sources in M\,82 have a WDB origin with relativistic particle efficiency comparable to values recently determined for galactic SNRs, but with higher (equipartition) magnetic fields, then a basis exists for the high ratio of radio luminosity to mechanical input energy for the WDB sources compared to galactic SNRs. {\it Compared to Galactic SNRs, the higher ratio of radio luminosity to available particle energy in WDBs in M\,82 is attributable to its strong interstellar field (1--2 orders stronger than the Galactic field), further amplified by the shocks associated with the expanding shells of the WDBs}. 

\subsection{Relativistic particle efficiency in the WDB model}

In \S3.1, the ratio ${\rm f = E_1/E_w}$ was set to f = 0.1 for the conservation of energy test. Here we further justify this choice by showing that the associated electron acceleration efficiency is comparable to that estimated for Galactic SNRs.

The efficiency of converting wind mechanical energy to relativistic particle energy is given by:
\begin{equation}
{\rm E_p/E_w = \frac{1}{2}\,f\,(1+k)^{4/7}\Phi^{3/7}},
\end{equation}
where the factor $\frac{1}{2}$ accounts for equal amounts of E$_1$ going into particle and magnetic energy. Since the electron energy is given by E$_{\rm p}$/(1+k), the efficiency of conversion to relativistic electron energy $\eta_{\rm e}$ is
\begin{equation}
{\rm \eta_e = \frac{1}{2}\,f\,[\Phi/(1+k)]^{3/7}}
\end{equation}
Setting f = 0.1 and adopting ${\rm 1\,\leqslant\,(1+k)\,\leqslant\,10}$, we obtain ${\rm 0.02\Phi^{3/7}\,\leqslant\,\eta_e\,\leqslant\,0.05\,\Phi^{3/7}}$, comparable to that in at least some Galactic SNRs. Thus f = 0.1 is seen to be a reasonable choice based on comparisons with recent results for SNRs. 

\subsection{Expansion, ionization and bremsstrahlung losses}

No account was taken of expansion, ionization or bremsstrahlung losses by the relativistic particles in the tests described in \S\,3, chiefly because they are less certain and/or less stringent. 
Expansion losses may be incurred as the particles propagate outward from the inner bubble shock toward the expanding outer shock, where the observed radiation is assumed to occur. Adiabatic losses scale inversely with ${\rm (V_2/V_1)^{1/3}}$ where V$_1$ and V$_2$ are the initial and final volumes occupied by the particles. We examine the worst case in which V$_1$ and V$_2$ constitute the entire volumes bounded by the inner wind and outer shock radii R$_{\rm sw}$ and R$_{\rm s}$, respectively. These radii vary as ${\rm R_{sw} \sim t^{2/5}}$ and ${\rm R_s \sim t^{3/5}}$, as given by equations 4.2 and 4.4 of \citet{koo}. For conditions relevant to this paper, these equations yield a current value of ${\rm (R_{sw}/R_s)\sim1/4 }$.
Using these expansion laws, assuming a steady injection of particles into the bubble at fixed fractional radius R$_{\rm sw}$, and integrating over R$_{\rm sw}$ since t=0, we estimate that the total energy loss is about 80\%\, for a cluster with a current value of ${\rm R_{sw}/R_s \approx 1/4}$. However, we emphasize here that this is a pessimistic estimate, and also a quite uncertain one especially since the geometrical factors adopted here would be quite different when the source of the wind has a finite radius.

Relativistic electrons co-existing with gas in the dense shell will suffer collisional (i.e. ionization) losses. Using equation 4.411 in \citet{lang} for the ionization loss, it may be shown that electrons emitting at frequency $\nu_5$ ($\nu$/5\,GHz) in a medium with neutral density 
n$_4$ (n/10$^4$\,cm$^{-3}$) and magnetic field H$_{-3}$ (H/10$^{-3}$\, gauss) lose energy on a time scale
\begin{equation} 
{\rm T_I = 1.0\,\times\,10^4\,n_4^{-1}\,H_{-3}^{-1/2}\,\nu_5^{1/2}\,   yr}.
\end{equation}    
Adopting n$_4$ = n$_s$ = 3.9, H$_{-3}$ = 1.5 for the shell, as suggested in \S 5.2.2, then ${\rm T_I\,\sim\,2\,\times\,10^3}$\, years for electrons radiating at 5\,GHz, which is less than the median age of the bubbles. Because of the energy (or frequency) dependence of this timescale, the effects of such losses would be most severe below 5\,GHz, and would produce a spectral turnover or cutoff at low frequencies. In fact, a turnover occurs below 1\,GHz in most sources in M\,82, but is attributed to the effects of free-free absorption by ionized gas in the ISM.

 A similar argument for the collision loss can be made if the shell gas is ionized, and the collision loss timescale (computation not shown here) in this case is estimated to be about a factor two shorter than the value for neutral gas. 

If the shell gas is ionized, then the relativistic electrons will also lose energy by bremsstrahlung radiation. From equation (4.417) of \citet{lang}, it may be shown that the loss time is only logarithmically dependent on particle energy (or frequency), and is given approximately by
\begin{equation}
{\rm T_B =  5.4\,\times\,10^3\,n_4^{-1}\,  yr}
\end{equation}
for electrons radiating near 5\,GHz in a field of  $1.5\,\times\,10^{-3}$\,gauss.

Again adopting n$_4$ = n$_s$ = 3.9, ${\rm T_B = 1.4\,\times\,10^3}$\,yr, once again less than the median age of the bubbles.  Thus it appears that ionization and/or bremsstrahlung losses would be significant if the relativistic particles were confined to the dense shell. However, it seems unlikely that the relativistic particles would be confined strictly to the dense shell, but would occupy a somewhat larger volume, likely encompassing the thin shell.

\section{THE SIZES OF CNRs AND SSCs}

The diameters of the CNRs in the output sample range from about 0.6 to 4.1\,pc. The mean photometric diameter for SSCs in M\,82 clusters observed by \citet{melo} is 
${\rm d = 11.4\,\pm\,2.8}$\,pc, largely independent of cluster mass. This difference clearly raises questions about the WDB picture since the clusters ought to be smaller than the bubbles they produce. However there are two considerations which could in principle resolve this difficulty. One is a possibility that mass segregation has occurred, with the most massive stars now at the center of the cluster. Although some observational evidence exists for this from color gradients \citep{mccrady}, the matter is unresolved, since there is also contradictory evidence \citep{smith}. Secondly, the hypothesized clusters underlying the CNRs are extremely young, since their winds are $\sim$\,10$^4$\,years old. Such clusters may be more compact than older clusters, which are likely modified by tidal interactions. Applying the Virial Theorem to a 10$^4$\,\msun\, cluster with a velocity dispersion of 10\,\kms, comparable to gas turbulent velocities within molecular clouds, yields d\,$\approx$\,0.5\,pc, smaller than those observed by more than an order of magnitude, and smaller than the CNR diameters. Thus a possible explanation is that the youngest clusters have large binding energies, which decrease with age due to energy input from tidal interactions with molecular clouds and other clusters.  

\section{THE CNRs IN OTHER WAVEBANDS}

M\,82 has been mapped in a number of wavebands from the radio to X-Rays, and such images could in principle yield evidence bearing on the relevance of the WDB origin for the CNRs. 

\subsection{Near-IR}

\citet{alonso} compared HST images in the [FeII] 1.644\,\micron\, line and adjacent continuum with the locations of 44 CNRs at sub arcsec resolution. The [FeII] lines are collisionally excited, and are well known signatures of SNRs both in our Galaxy and nearby galaxies. They find that 30\,--\,50\% of the CNRs possess associated line emission, which they consider support for the SNR interpretation. 
The observed luminosities (corrected for extinction) are in the range $3\times 10^{36}-2.2\times 10^{38}$\,erg\,s$^{-1}$.

For comparison, we have estimated the [FeII] luminosity expected for a typical WDB shell. There are two possible sources - one with luminosity L$_{\rm s}$, associated with the outer shock, and the other with luminosity L$_{\rm p}$, associated with gas in the shell photoionized by the cluster. It is shown in the Appendix that these may be written as

\begin{equation}
{\rm L_s = 2.0\,\times\,10^{37}\,\eta_s\,M_4\, erg\,s^{-1}}
\end{equation}
\begin{equation}
{\rm L_p = 3.3\,\times\,10^{37}\,\eta_p\,M_4\, erg\,s^{-1}}
\end{equation}

where ${\rm \eta_s}$ and ${\rm \eta_p}$ are the ratios of [FeII] luminosity to shock kinetic energy and Paschen$\beta$ luminosity, respectively. \citet{mouri} estimate ${\rm \eta_s=2\,\times\,10^{-3}}$ for M\,82. For ${\rm \eta_p}$, their Figure\,8(a) shows a plot of ${\rm \eta_p}$ vs.\,ionization parameter U for a limited set of conditions. In our case ${\rm U=Ly/4\,\pi\,r^2\,n_s\,c}$, where c is the speed of light. Assuming solar abundances, ${\rm n_s=10^4}$\,cm$^{-3}$, and ionization by a blackbody with T=40,000\,K, then for ${\rm M_4=2.0}$, and ${\rm r=1}$\,pc, we estimate ${\rm \eta_p=0.3}$. The resulting luminosities are ${\rm L_s=8.0\,\times\,10^{34}}$\,erg\,s$^{-1}$ and ${\rm L_p=2.0\,\times\,10^{37}}$\,erg\,s$^{-1}$, showing that the emission from the photoionized shell dominates that from the shock by more than two orders of magnitude. Comparison with the observed luminosity range above shows that the [FeII] luminosity of a typical WDB would be in the middle of the observed range.

The ratio ${\rm \eta_p}$ may also be used to predict the ratio [FeII]/Br$_\gamma$ for comparison with the observed ratios in Figure 10 of \citet{alonso}. The predicted ratio is about 0.4, compared to a median observed value near 1.0. Both are less than that shown for Galactic SNRs which is about 10. The predicted values of L$_{\rm p}$ and [FeII]/Br$_\gamma$ would be higher for lower ionization parameter, for e.g.\,if the shell density were as high as our estimate of $4\,\times\,10^4$\,cm$^{-3}$ discussed in section 5.2.2. Higher densities still would not help however since this value is approximately the critical density for this transition, setting a corresponding limit on L$_{\rm p}$.

It is also shown in \citet{mouri} that if the abundances are sub-solar, with 90\%\, of Fe locked up in grains, then ${\rm \eta_p}$ and hence L$_{\rm p}$ would be lower than the above estimate by an order of magnitude, and a typical WDB cluster probably would not have been detected by \citet{alonso}.
 
\citet{alonso} did not conduct a search for coincidences between CNRs and continuum near-IR sources. Their Figure 2(a) shows the image in the adjacent continuum band at 1.64\,\micron. Using a position and flux calibrated electronic version of this image provided by the authors, we searched for coincidences between 1.64\,\micron\, continuum point sources and both CNRs and thermal compact sources. For the CNRs two coincidences were found with separation $\leqslant$\,0.5\arcsec, which is consistent with chance associations (estimated probability 20\,--\,30\%). For the thermal sources there were two coincidences with separation $\leqslant$\,0.2\arcsec , which may also be chance associations. Thus there is no significant indication that the near-IR continuum and radio populations are physically associated. 

We estimate that the faintest detectable continuum point source on these images is about 0.3\,mJy. For comparison, SB99 predicts that, with cluster parameters used in this paper, an expected 1.64\micron\, continuum flux of 0.12\,M$_4$\,mJy from the cluster stars.  With an estimate of  typically 2$^m$ of extinction at H (based on a typical value of 1.4$^m$ at K-band estimated by Alonso-Herrero et al.), we conclude that if there are clusters associated with compact radio sources, they would likely be below the detection threshold of these images. 

\subsection{Mid-IR}

IR emission at wavelengths near 10\,\micron\, is expected in the WDB model from dust in the shells heated by optical radiation from the underlying clusters. The shells swept up by the shock are nearly opaque in the visual region as shown in \S\,5.2.1, so it may be assumed for present purposes that all of the bolometric luminosity is converted to IR emission. Recent high resolution (0.4\arcsec) observations of the central 400\,pc of M\,82 by \citet{lipscy} at 11.7 and 17.65\,\micron\, reveal seven star-forming clusters which together provide $\sim$\,15\% of the total mid-IR luminosity. The authors kindly provided electronic versions of these images to permit a search for associations with the CNRs. A search at both wavelengths was made and failed to yield any coincidences with the output sample members. In general, the upper limits for coincident sources were about 50\,mJy and 100\,mJy at 11.7 and 17.65\,\micron\, respectively. 

A crude and optimistic estimate of the expected level of mid-IR flux was made by assuming, as in the case of the Milky Way, that small grains absorb 20\,--\,30\% of the incident optical light and re-radiate this energy in the mid-IR as a result of impulsive grain heating \citep{draine}. The remaining optical light is radiated in the far-IR by larger grains under equilibrium conditions. By assuming that the mid-IR emission acts like a black body with peak emission at 12\,\micron\, and total luminosity equivalent to 30\%\, of the bolometric luminosity given by equation (3d), we obtain 44\,mJy for the flux associated with the $2\,\times\,10^4$\,\msun\,cluster, just below the mean upper limit of 50\,mJy at 11.7\,\micron.  This result suggests that such clusters might be detectable if the sensitivity were improved by a factor of several. 

\subsection{X-Ray emission}

Emission at soft X-Rays is expected from the hot expanding bubble. High angular resolution observations of M\,82 using Chandra show about 25 discrete sources down to a luminosity detection limit of $2\,\times\,10^{37}$\,erg\,s$^{-1}$ \citep{zezas}. Apart from the powerful variable source tentatively identified with the CNR 41.30\,+\,59.6, these sources appear in the majority to be X-Ray binaries associated with young star clusters, and not identified with CNRs. 

To estimate the expected X-Ray luminosity of a typical CNR in the WDB picture, we first note that the total wind power for a $2\,\times\,10^4$\,\msun\, cluster is $1.1\,\times\,10^{38}$\,erg\,s$^{-1}$ which represents an upper limit to the X-ray luminosity of the bubble. The true X-Ray luminosity would be only a fraction of this amount because: (1) the thermal energy of the bubble is 5/11 that of the total wind mechanical energy \citep{maclow}; (2) the bubble is in the adiabatic expansion phase and the cooling rate at this stage is lower than the wind power; and (3) most of the cooling radiation is emitted in the interface region between the bubble shock and outer shell by UV fluorescence lines such as OVI\, $\lambda$\,1035 \citep{weaver}. Consequently, the X-ray luminosity of a typical bubble could be weaker than the current detection limit.

\subsection{Association with molecular clouds}

Using CO J=2--1 interferometric observations of \object{M\,82} at sub-arcsec resolution, \citet{keto} resolved the molecular gas in the starburst region into a large number of compact clouds with masses ranging from $2\,\times\,10^3$ to $2\,\times\,10^6$\,\msun. Surprisingly, the CNRs seem to cluster around the periphery of molecular emission regions, and there is almost an anti-correlation between molecular clouds and the locations of the compact radio sources. The authors suggest that, since the CNRs must be signposts of recent star formation, the associated gas has either been fully consumed by stars, or any associated gas has a spatial scale $>$\,10\arcsec. Perhaps it is noteworthy that in \S\,4 it was shown that in the WDB model, the ISM into which the bubbles are expanding is about $2.6\,\times\,10^3$\,cm$^{-3}$, comparable to that estimated for the warm diffuse inter-clump medium ($\sim$\,10$^3$\,cm$^{-3}$) rather than the cooler and denser clumps ($\sim$\,10$^5$\,cm$^{-3}$) where stars form. Thus the clusters could have consumed the dense gas, and are now expanding into a remnant warm, primarily dissociated medium, invisible in CO emission. 

\section{SUMMARY AND CONCLUSIONS}

We have introduced and tested the hypothesis that many of the CNRs in M\,82 are associated with WDBs rather than SNRs. A primary motivation for this hypothesis is that most of the radio sources exhibit no discernible time variation in their flux densities, suggesting an age greater than expected for SNRs, but comfortably within the range expected for WDBs.  In the WDB picture, the synchrotron emitting particles are accelerated by the first order Fermi mechanism at the shock at the site of the interaction between the cluster wind and bubble gas. 

The WDB model was tested using a limited but representative sample of objects with well defined radio spectra. The model, based on the theory of WDBs and model parameters from SB99, was constrained to satisfy the energy requirements of the synchrotron sources and limits imposed by losses of the synchrotron emitting particles and radiative cooling time scale of the bubble. Further constraints were that the predicted free-free flux density from the HII region produced by the young cluster not exceed the upper limits set by observation, and that the bubble age not exceed the time to pressure confinement. These considerations constrain the associated cluster masses to $\sim2\times10^{4(\pm\,1)}$\,\msun\,and the surrounding ISM density to $\sim3\times10^{3(\pm\,1)}$\,cm$^{-3}$. The corresponding ages cover the range from a few $\times\,10^3$\, years to a few $\times\,10^4$\, yr. This range in age is shown to be consistent with a plausible star formation rate in M\,82. Estimates of the flux densities in various wavebands of clusters at the median mass of $2\,\times\,10^4$\,\msun\, show that published observations are not yet sensitive enough to confirm (or rule out) such clusters, but in some cases are very close.

Thus, we conclude that the WDB hypothesis appears to be a plausible alternative to the SNR picture for the origin of the CNRs, but the tests do not rule out an SNR origin. It is also possible that the CNRs comprise a mixture of both types. It is clear that some critical tests are needed to discriminate between the two ideas. The most stringent tests will be: (1) a search for free-free emission from ionized shells surrounding these clusters at mm wavelengths, permitting a clear discrimination between thermal and non-thermal emission; (2) a search for near-IR emission with colors signifying the stellar content of the SSC. Such continuum emission would not be espected from the remnant of an isolated SN since any associated cluster would produce a SN wind driven supershell rather than an isolated remnant; (3) near IR measurement of the ratios of [FeII] to recombination line emission; and (4) measurement of shell expansion velocities. These tests are feasible now with deep searches and radio observations at high angular resolution made over a long time baseline.

The WDB picture needs to be harmonized with the larger picture of M\,82. For example, why aren't young SSCs with masses outside the range indicated in Figures 2 and 3 also associated with non-thermal sources? In the WDB model, should we see an association between optically visible SSCs and CNRs? If the CNRs are not SNRs then where are the latter objects? We consider these issues in turn.

Naturally, SSCs with masses lower than the observed range would not have winds powerful enough to yield sufficient relativistic particle energy. Clusters with higher masses could emit synchrotron radiation, but such emission could be suppressed by inverse Compton cooling if the ambient gas density is high enough to confine the bubble for a sufficiently long period, as shown in section 5.1.

Regarding the potential association between CNRs and visible SSCs, such identification is unlikely, since these clusters are found to have ages $>10^6$\,yr \citep{melo}, and are thus too old to exhibit nonthermal emitting shells. The very young clusters associated with this CNR phenomenon would very likely be obscured by the surrounding dust in the molecular clouds where they are formed. In addition, the current lack of astrometric precision for optically visible clusters precludes a serious search for identification, especially for the faint members with lower masses (near $10^4$\,\msun).

Concerning the SNRs in \object{M\,82}, SNe may originate predominantly in SSCs if such clusters are the most common sites of massive star formation. Individual SNRs may then be more rare than previously assumed, since the aggregate effect of SNe ejecta in clusters is to produce SN winds even more powerful than the earlier stellar winds. The SN winds would re-energize the pressure confined shells produced by the earlier winds to produce the supershells evident in \object{M\,82} by free-free and HI absorption of the non-thermal background and by CO emission from molecular gas \citep{wills,weiss,pedlar}. Though such shells may also produce non-thermal emission, their size (tens to hundreds of pc) would preclude easy identification because of the general confusion by the complex non-thermal background. It is undoubtedly such shells, particularly from massive clusters, which break out of the disk and contribute to energizing the nuclear outflows in \object{M\,82} and other starburst galaxies. Note that if many of the CNRs are not SNRs, then the SNe rate based on CNRs alone would be overestimated. However, this deficit may be more than compensated by the complementary rates inferred from the interpretation that supershells are the product of multiple SNe in SSCs.

The compact non-thermal sources in \object{M\,82} and other galaxies may thus be the unique signatures of the winds of low mass very young clusters embedded in the ISM, except for some individual cases where their time variability and/or measured expansion rate identify them uniquely as SNRs.

\acknowledgments

This research was supported by a Discovery Grant to ERS from the Natural and Engineering Sciences Council of Canada and financial support to MS from the University of Toronto. We thank A. Alonso-Herrero and S.J. Lipscy for providing electronic versions of their published near-IR and mid-IR images respectively. We also thank an anonymous referee for helping us to improve the manuscript. 

\appendix

\section{Appendix: [FeII] Emission from Cluster Wind Driven Bubbles}

There are two potential sources of [FeII] emission from a WDB. The first is collisionally excited gas in the radiative shock formed as the interstellar medium is swept up into a shell by the expanding bubble. The second is the part of this shell photoionzed by the cluster. \citet{mouri} has modeled [FeII] 1.257\,\micron\, emission from both shocked and photoionized gas, and we base our estimate on this work. The results are applicable as well to emission at 1.644\,\micron, relevant to the discussion in \S\,8.1, since both transitions have a common upper state and have very similar transition probabilites \citep[e.g.][]{nussbaumer}.

(a) Emission from shocked gas in the shell

The [FeII] luminosity L$_s$ from the shocked gas may be parameterized as a fraction ${\rm \eta_s=L_s}$/\ek\, of the flow rate of kinetic energy \ek\, of ambient gas into the shock, which is also the rate of energy radiated by the shell. The parameter ${\rm \eta_s}$ is given in Figure 11(b) of \citet{mouri}. The quantity \ek\, is given by \ek$=(27/77)$\,L$_{\rm w}$ \citep{maclow}. Combining this with equation (3b) gives

\begin{equation}
{\rm L_s = 2.0\,\times\,10^{37}\,\eta_s\,M_4\, erg\,s^{-1}}
\end{equation}

(b) Emission from photoionized gas in the shell

The [FeII] luminosity L$_p$ is estimated here for an ionization bounded shell, and may be written as ${\rm L_p=(L_p/Pa\beta)(Pa\beta/Ly)Ly}$, where Pa$\beta$ and Ly are respectively the Paschen $\beta$ and Lyman continuum luminosities. The ratio ${\rm \eta_p=(L_p/Pa\beta)}$ is dependent on the assumed metal abundances, the shell gas density, the spectrum of the source of ionization, and the ionization parameter. It is given by Figure 8(a) of \citet{mouri} for a limited set of conditions. The ratio (Pa$\beta$/Ly) may be computed from standard case B recombination \citep[e.g.][]{osterbrock} using an electron temperature of 10$^4$\,K. Combining this result with equation (3e) then gives

 \begin{equation}
{\rm L_p = 3.3\,\times\,10^{37}\,\eta_p\,M_4\, erg\,s^{-1}}
\end{equation}


\end{document}

%% file: tab1.tex
\begin{table}
\begin{center}
\caption{Physical parameters, derived SSC masses, and interstellar medium densities for an acceptable fit of the CNRs to the WDB model for the output sample}

\begin{tabular}{cccccccc}
\tableline\tableline
Name & \multicolumn{1}{c}{d\tablenotemark{a}} & \multicolumn{1}{c}{S$_0$\tablenotemark{b}} &  \multicolumn{1}{c}{$\alpha$\tablenotemark{c}}  &    \multicolumn{1}{c}{E$_1$\tablenotemark{d}}  &  \multicolumn{1}{c}{H$_1$\tablenotemark{e}}  & \multicolumn{1}{c}{$log$\,M\tablenotemark{f}}  & 
\multicolumn{1}{c}{$log$\,n$_0$\tablenotemark{g}} \\
  & pc  & mJy  &  & 10$^{48}$\,erg & 10$^{-4}$\,gauss  & (\msun) & (cm$^{-3}$)\\
\tableline
 & & & & & & (median, range) & (median, range)\\
\tableline
39.1+57.4 & 1.7 & 8.6 & -0.38 & 3.3 & 7.6 &  4.3 (4.0, 4.5) & 3.4 (2.5, 4.0) \\
39.4+56.1 & 3.3 & 3.1 & -0.21 & 5.0 & 3.4 &  4.2 (3.5, 4.5) & 2.9 (1.5, 4.0) \\
39.6+53.4 & 2.8 & 4.4 & -0.71 & 4.8 & 4.3 &  4.0 (4.0, 4.0) & 3.5 (3.0, 4.0) \\
39.8+56.9 & 3.1\tablenotemark{h} & 2.1 & -0.49 & 3.4 & 3.0 & 4.2 (3.5, 4.5) & 2.7 (1.0, 4.0) \\
40.3+55.1 & 1.6 & 2.3 & -0.55 & 1.4 & 5.5 &  4.2 (3.5, 4.5) & 3.1 (1.5, 4.0) \\
40.6+56.1 & 2.9 & 3.8 & -0.72 & 4.7 & 3.9 &  3.9 (3.5, 4.0) & 3.3 (2.5, 4.0) \\
40.7+55.1 & 2.0 & 7.9 & -0.52 & 5.9 & 8.0 &  4.5 (4.4, 4.5) & 3.2 (3.0, 3.5) \\
41.3+59.6 & 1.1 & 8.6 & -0.54 & 2.0 & 11.1&  4.3 (4.4, 4.5) & 3.4 (2.5, 4.0) \\
42.7+55.7 & 4.1 & 4.4 & -0.61 & 7.1 & 2.9 &  3.5 (3.5, 3.5) & 3.5 (3.5, 3.5) \\
43.2+58.4 & 1.0 & 15.3 & -0.67 & 2.6 & 14.5 & 4.5 (4.5, 4.5) & 3.7 (3.5, 4.0) \\
43.3+59.2 & 0.6 & 30.3 & -0.64 & 1.9 & 28.0 & 4.5 (4.5, 4.5) & 4.0 (4.0, 4.0) \\
44.0+59.6 & 0.8 & 62.0 & -0.51 & 3.8 & 25.6 & 5.0 (5.0, 5.0) & 3.5 (3.5, 3.5) \\
44.3+59.3 & 1.9 & 6.7 & -0.56 & 3.3 & 6.3 & 4.3 (4.0, 4.5) & 3.2 (2.0, 4.0) \\
44.5+58.2 & 2.2 & 7.2 & -0.61 & 4.2 & 5.8 & 4.0 (4.0, 4.0) & 3.5 (3.0, 4.0) \\
44.9+61.1 & 1.6 & 6.6 & -0.45 & 2.6 & 7.3 & 4.0 (4.0, 4.0) & 3.7 (3.5, 4.0) \\
45.2+61.2 & 1.1 & 24.1 & -0.68 & 4.1 & 15.4 & 4.5 (4.5, 4.5) & 3.5 (3.5, 3.5) \\
45.3+65.2 & 2.0 & 5.4 & -0.62 & 3.1 & 5.7 & 3.9 (3.5, 4.0) & 3.6 (3.0, 4.0) \\
45.4+67.4 & 2.2 & 4.7 & -0.57 & 3.5 & 5.2 & 4.0 (4.0, 4.0) & 3.5 (3.0, 4.0) \\
45.5+64.8 & 3.1\tablenotemark{h} & 1.7 & -0.15 & 3.8 & 3.2 & 4.2 (3.5, 4.5) & 3.0 (1.5, 4.0) \\
45.8+65.3 & 2.1 & 5.1 & -0.23 & 3.8 & 5.8 & 4.3 (4.0, 4.5) & 3.2 (2.0, 4.0) \\
45.9+63.9 & 2.2 & 4.1 & -0.53 & 3.2 & 4.9 & 4.3 (4.0, 4.5) & 3.2 (2.0, 4.0) \\
46.5+63.9 & 1.5 & 5.2 & -0.57 & 2.0 & 7.3 & 3.9 (3.5, 4.0) & 3.6 (3.0, 4.0) \\
46.6+73.8 & 3.1\tablenotemark{h} &  3.7 & -0.78 & 5.8 & 3.9 & 4.0 (4.0, 4.0) & 3.2 (3.0, 3.5) \\
46.7+67.0 & 3.0 & 5.2 & -0.57 & 5.0 & 4.0 & 4.3 (4.0, 4.5) & 3.2 (2.0, 4.0) \\
47.4+68.0 & 3.1\tablenotemark{h} & 4.1 & -0.83 & 6.6 & 4.2 & 4.3 (4.0, 4.5) & 2.9 (2.0, 3.5) \\
\tableline
\end{tabular}
\tablenotetext{a}{source diameter}
\tablenotetext{b}{flux at 1\,GHz}
\tablenotetext{c}{spectral index (${\rm S=S_0\,\nu^{\alpha}}$)}
\tablenotetext{d}{equipartition energy = ${\rm E_1\times\,(1+k)^{4/7}\,\phi^{3/7}}$}
\tablenotetext{e}{equipartition magnetic field = ${\rm H_1\times\,[(1+k)/\phi]^{2/7}}$}
\tablenotetext{f}{SSC mass (median and range)}
\tablenotetext{g}{ISM hydrogen density (median and range)}
\tablenotetext{h}{based on assumed diameter of 0.2\arcsec}
\tablecomments{Source data obtained from \citet{huang}, \citet{muxlow}, \citet{allen} and \citet{mcdonald}} 
\end{center}
\end{table}

%% file: ms.bbl
\begin{thebibliography}{}

\bibitem[Allen(1999)]{allen} Allen, M. L. 1999, Ph.D. Thesis
\bibitem[Alonso-Herrero et al.(2003)]{alonso} Alonso-Herrero, A., Rieke, G. H., Rieke, M. J., \& Kelly, D. M. 2003, \aj, 125, 1210
\bibitem[Berezhko\,\&\,V\"olk(1997)]{berezhko} Berezhko, E. G. \& V\"olk, H. J. 1997, APh, 7, 183 
\bibitem[Beswick et al.(2006)]{beswick} Beswick, R. J., Riley, J. D., Marti-Vidal, I., Pedlar, A., Muxlow, T. W. B., McDonald, A. R., Wills, K. A., Fenech, D., \& Argo, M. K. 2006, \mnras, 369, 1221
\bibitem[Blandford\,\&\,Eichler(1987)]{blandford} Blandford, R. \& Eichler, D. 1987, \physrep, 154, 1 
\bibitem[Bohlin, Savage\,\&\,Drake(1978)]{bohlin} Bohlin, R. C., Savage, B. D., \& Drake, J. F. 1978, \apj, 224, 132
\bibitem[Chevalier\,\&\,Fransson(2001)]{chevalier} Chevalier, R. A. \& Fransson, C. 2001, \apjl, 558, L27
\bibitem[Condon(1992)]{condon} Condon, J. J. 1992, \araa, 30, 575
\bibitem[Draine(2003)]{draine} Draine, B. T. 2003, \araa, 41, 241
\bibitem[Drury(1983)]{drury} Drury, L. O'C, 1983, Rep. Prog. Phys. 46, 973
\bibitem[Dyer et al.(2001)]{dyer} Dyer, K. K., Reynold, S. P., Borkowski, K. J., Allen, G. E., \& Petre, R. 2001, \apj, 551, 439
\bibitem[Ellison et al.(2004)]{ellison} Ellison, D. C., Decourchelle, A., \& Ballet, J. 2004, \aap, 413, 189
\bibitem[Fenech et al.(2005)]{fenech} Fenech, D., Muxlow, T. W. B., Pedlar, A., Beswick, R., Argo, M. K., \& Wills, K. 2005, \memsai, vol. 76, 583
\bibitem[Fields et al.(2001)]{fields} Fields, B. D., Olive, K. A., Cass\'e, M., \& Vangioni-Flam 2001, \aap, 370, 623
\bibitem[F\"orster Schreiber et al.(2003)]{forster} F\"orster Schreiber, N. M., Genzel, R., Lutz, D., \& Sternberg, A. 2003, \apj, 599, 193
\bibitem[Ginzburg\,\&\,Syrovatskii(1964)]{ginzburg} Ginzburg, V. L. \& Syrovatskii, S. I. 1964, The Origin of Cosmic Rays (New York: Pergamon Press, The MacMillan Company)
\bibitem[G\"usten et al.(1993)]{gusten} G\"usten, R., Serabyn, E., Kasseman, C., Schinkel, A., Schneider, G., Schulz, A., \& Young, K. 1993, \apj, 402, 537
\bibitem[Huang et al.(1994)]{huang} Huang, Z. P., Thuan, T. X., Chevalier, R. A., Condon, J. J., \& Yin, Q. F. 1994, \apj, 424, 114
\bibitem[Keshet et al.(2003)]{keshet} Keshet, U., Waxman, E., Loeb, A., Springel, V., \& Hernquist, L. 2003, \apj, 585, 128
\bibitem[Keto, Ho\,\&\,Lo(2005)]{keto} Keto, E., Ho, L. C., \& Lo, K.-Y. 2005, \apj, 635, 1062
\bibitem[Klein, Wielebinski\,\&\,Morsi(1988)]{klein} Klein, U., Wielebinski, R., \& Morsi, H. W. 1988, \aap, 190, 41
\bibitem[Koo\,\&\,McKee(1992)]{koo} Koo, B.-C. \& McKee, C. F. 1992, \apj, 388,93
\bibitem[Kronberg et al.(2000)]{kronberg} Kronberg, P. P., Sramek, R. A., Birk, G. T., Dufton, Q. W., Clarke, T. E., \& Allen, M. L. 2000, \apj, 535, 706
\bibitem[Lang(1998)]{lang} Lang, K. 1998, in Astrophysical Formulae vol. 1, Radiation, Gas Processes and High Energy Astrophysics, third edition, Springer-Verlag: Berlin, Heidelberg, New York
\bibitem[Larking et al.(1994)]{larkin} Larkin, J. E., Graham, J. R., Matthews, K., Soifer, B. T., Beckwith, S., Herbst, T. M., \& Quillen, A. C. 1994, \apj, 420, 159
\bibitem[Leitherer et al.(1999)]{leitherer} Leitherer, C., Schaerer, D., Goldader, J. D., Delgado, R. M. G., Robert, C., Kune, D. F., de Mello, D., Devost, D., \& Heckman, T. M. 1999, \apjs, 123, 3
\bibitem[Lipscy\,\&\,Plavchan(2004)]{lipscy} Lipscy, S.J. \& Plavchan, C. 2004, \apj, 603, 82
\bibitem[Mac Low\,\&\,McCray(1988)]{maclow} Mac Low, M.-M. \& McCray, R. 1988, \apj, 324, 776
\bibitem[Mao et al.(2000)]{mao} Mao, R .Q., Henkel, C., Schulz, A., Zielinski, M., Mauersberger, R., St\"orzer, H., Wilson, T.L., \& Gensheiner, P. 2000, \aap, 358, 433
\bibitem[McDonald et al.(2002)]{mcdonald} McDonald, A. R., Muxlow, T. W. B., Wills, K. A., Pedlar, A., \& Beswick, R. J. 2002, \mnras, 334, 912
\bibitem[McCrady, Graham\,\&Vacca(2005)]{mccrady} McCrady, N., Graham, J. R., \& Vacca, W. D. 2005, \apj, 621, 278
\bibitem[McCray\,\&\,Kafatos(1987)]{mccray} McCray, R. \& Kafatos, M. 1987, \apj, 317, 190
\bibitem[Melo et al.(2005)]{melo} Melo, V. P., Mu\~noz-Tu\~n\'on, C., Ma\'iz-Appel\'aniz, J., \& Tenorio-Tagle, G. 2005, \apj, 619, 270
\bibitem[Mouri et al.(2000)]{mouri} Mouri, H., Kawara, K. \& Taniguchi, Y. 2000, \apj, 528, 186 
\bibitem[Muxlow et al.(1994)]{muxlow} Muxlow, T. W. B., Pedlar, A., Wilkinson, P. N., Axon, D. J., Sanders, E. M., \& de Bruyn, A. G. 1994, \mnras, 266, 455
\bibitem[Muxlow et al.(2005)]{muxlow05} Muxlow, T. W. B., Pedlar, A., Beswick, R. J., Argo, M. K., O'Brien, T. J., Fenech, D. \& Trotman, W. 2005, \memsai, vol. 76, 516
\bibitem[Nussbaumer\,\&\,Storey(1988)]{nussbaumer} Nussbaumer, H. \& Storey, P. J. 1988, \aap, 193, 327
\bibitem[Oey et al.(2002)] {oey} Oey, M. S., Groves, B., Stavely-Smith, L., \& Smith, R. C. 2002, \aj, 23, 255 
\bibitem[Osterbrock(1989)]{osterbrock} Osterbrock, D. E. 1989, Astrophysics of Gaseous Nebulae, University Science Books, Mill Valley, California
\bibitem[Pacholczyk(1970)]{pacholczyk} Pacholczyk, A. 1970, Radio Astrophysics (San Francisco, CA: Freeman)
\bibitem[Pedlar, Muxlow\,\&\,Wills(2003)]{pedlar} Pedlar, A., Muxlow, T., \& Wills, K. 2003, in Winds, Bubbles and Explosions: A Conference to Honour John Dyson, eds. S.J. Arthur \& W.J. Henney, RevMexAA (Serie de Conferncias), 15, 303
\bibitem[Pittard et al.(2006)]{pittard} Pittard, J. M., Dougherty, S. M., Coker, R. F., O'Connor, E., \& Bolingbrooke, N. J. 2006, \aap, 446, 1001
\bibitem[Rieke et al.(1980)]{rieke} Rieke, G. ., Lebofsky, M. ., Thompson, R. ., Low, F. J., \& Tokunaga, A. T. 1980, \apj, 238, 24
\bibitem[Rodriguez-Rico et al.(2004)]{rrico} Rodriguez-Rico, C. A., Viallefond, F., Zhao, J.-H., Goss, W. M., \& Anantharamaia, K. R. 2004, \apj, 616, 783
\bibitem[Smith et al.(2006)]{smith} Smith, L. J., Westmoquette, M. S., Gallagher, J. S., O'Connell, R. W., Rosario, D. J., \& de Grijs, R. 2006, \mnras, 370, 513
\bibitem[Thompson et al.(2006)]{thompson} Thompson, T. A., Quataert, E., Waxman, E., Murray, N. \& Martin, C. L. 2006, \apj, 645, 186
\bibitem[Van Loo, Runacres\,\&\,Blomme(2006)]{vanloo} Van Loo, S., Runacres, M. C., \& Blomme, R. 2006, \aap, 452, 1011
\bibitem[Weaver et al.(1977)]{weaver} Weaver, R., McCray, R., Castor, J., Shapiro, P., \& Moore, R. 1977, \apj, 218, 377
\bibitem[Wei\ss\, et al.(1999)]{weiss} Wei\ss, A., Walter, F., Neininger, N., \& Klein, U. 1999, \aap, 345, 23
\bibitem[Wentzel (1974)]{wentzel} Wentzel, D. G. 1974, \araa, 12, 71
\bibitem[Wills et al.(1997)]{wills} Wills, K. A., Pedlar, A., Muxlow, T. W. B., \& Wilkinson, P. N. 1997, \mnras, 291, 517
\bibitem[Zezas et al.(2004)]{zezas} Zezas, A., Fabbiano, G., Ward, M., Schweizer, F., King, A., Kilgard, R., Kaaret, P., \& Prestwich, A. 2004, Proceedings of 35th COSPAR Scientific Assembly, 18-25 July 2004, Paris, France, p. 4107
\end{thebibliography}
